\documentclass[aps,prl,twocolumn,groupedaddress]{revtex4}
\usepackage{epsfig}
\usepackage{psfig}
\date{\today}

\newcommand{\cR}{{\mathcal R}}
\newcommand{\cS}{{\mathcal S}}
\newcommand{\cZ}{{\mathcal Z}}

\begin{document}
\title{Canonical-grandcanonical ensemble in-equivalence in Fermi systems?}
\author{Drago\c s-Victor Anghel}
\affiliation{University of Oslo, Department of Physics, P.O. Box 1048 - Blindern, N-0316 Oslo, Norway.}
\begin{abstract}\noindent
I discuss the effects of fermionic condensation in systems of 
constant density of states. I show that the condensation leads 
to a correction of the chemical potential and of the Fermi distribution 
in canonical Fermi systems at low temperatures. This implies that the 
canonical and grandcanonical ensembles are not equivalent even for 
Fermi systems. 
\end{abstract}
\maketitle

\section{Introduction} \label{reminder}

Let a system $\cS$ be in contact with a heat and particle reservoir, 
$\cR$. The microstates of the system will be denoted by $m_i$. If I 
assume that the system is ergodic 
and all the microstates corresponding to any fixed $E$ and $N$ are 
equally probable, then the probability associated to any state, say 
$p(m_{E,N})$ satisfies 
\begin{equation} \label{psEN}
p(m_{E,N}) \propto e^{\beta(E-\mu N)} \,, 
\end{equation}
where $\beta\equiv 1/(k_B T)$, $T$ is 
the temperature, and $\mu$ is the chemical potential of the reservoir. 

The same probability distribution may be obtained if we take as starting 
point information theory \cite{jaynes}. The information we start with is 
that the average energy and particle number are $E$ and $N$, respectively. 
Then, the probability distribution (\ref{psEN}) is the least biased 
estimate possible on the given information \cite{jaynes}. If the system 
is ergodic or not, is of no 
importance from this point of view. For a macroscopic 
system the maximum of the probability distribution is sharply peaked 
around the average values, $\langle N\rangle$ and $\langle E\rangle$, 
so both, particle number and internal energy, have well defined values. 

Now let us assume that we can calculate (and measure) a macroscopic 
parameter, which I shall denote by $X$. The probability distribution over 
the microstates is given by Eq. (\ref{psEN}) and from this one can calculate 
the probability distribution over the parameter $X$: 
\begin{equation} \label{PX}
P_X(X) = \sum_i^{X(m_i)=X} p(m_i) \,.
\end{equation}
If $X$ is well defined, then $P_X(X)$ should have a sharp peak at 
$X=\langle X\rangle$. If $P_X(X)$ has two maxims, then the system 
undergoes a phase transition. In each of the phases, one maximum 
dominates and this fixes the value of $X$ for that phase \cite{lee,FC1}. 

For a system of fermions 
a parameter which is 
surprisingly interesting to analyze is the number of particles that 
occupy completely an energy interval (no holes left in this interval), 
starting at the bottom of the single particle spectrum \cite{FC1,FC2,FC3} -- 
let me call this parameter $N_0$ and the energy interval $[0,\epsilon_0]$. 
In \cite{FC1} I gave an example of an interacting system for which the 
probability distribution $P_{N_0}(N_0)$ forms, below a certain temperature, 
two competing maxims. One maximum, which exists for any $T>0$, is located at 
$N_0=0$ and the other appears at finite $N_0$. At transition temperature 
the maximum centered at $N_0>0$ equals the maximum existent at 
$N_0=0$ and a first order phase transition occurs. Above transition 
temperature, since $P_{N_0}(N_0)$ is maximum at $N_0=0$, 
$\langle N_0\rangle\ (\gtrsim 0)$ 
is microscopical. Below transition temperature $P(N_0)$ is maximum 
at $N_0>0$, so $\langle N_0\rangle>0$ is a macroscopic quantity. 
Due to the interaction, an energy gap is formed between the 
degenerate $N_0$ particles and the rest of the particles. 

The same parameter may be analyzed for a system of ideal fermions. 
Assume that the density of states (DOS) has the general form 
$\sigma(\epsilon)=C\epsilon^s$, where $C$ and $s$ are constants. 
Now I require that $N_0$ particles form a degenerate subsystem on 
the  first $N_0$ energy levels, and the first hole in the spectrum 
appears at energy $\epsilon_0$ (or energy level $N_0+1$ -- see Ref. 
\cite{FC2} for details). Using 
again Eq. (\ref{PX}), I calculate $P_{N_0}(N_0)$ or 
$P_{\epsilon_0}(\epsilon_0)$. 
If $\cZ$ is the partition function of the system, and $\cZ_{N_0}$ 
is the number of configurations with the first hole appearing at 
$N_0+1$, then $P_{N_0}(N_0)=\cZ_{N_0}/\cZ$ and 
$P_{\epsilon_0}(\epsilon_0)=\sigma(\epsilon_0)\cdot P_{N_0}[N_0(\epsilon_0)]$. 
Since $\cZ$ is a constant, the extrema of $P$ are found by solving 
$d \cZ_{N_0}/d N_0=0$, or $d\cZ_{\epsilon_0}/d\epsilon_0=0$. Even more 
convenient is to work with $\log Z_{N_0}$, which is  \cite{FC2} 
\begin{eqnarray} 
\log\cZ_{N_0} &=& \left[ -\beta \left(C\frac{\epsilon_0^{s+2}}{s+2}-
\epsilon_0\right)+\beta\mu \left(C\frac{\epsilon_0^{s+1}}{s+1} -1
\right) \right] \nonumber \\
&&+ C\int_{\epsilon_0}^\infty d\epsilon\, \epsilon^s 
\log\left[1+e^{-\beta(\epsilon-\mu)} \right] \, .\label{logZN0}
\end{eqnarray}
Since $d\cZ_{N_0}/d N_0=(d\cZ_{N_0}/d\epsilon_0)\cdot(d\epsilon_0/dN_0)=(d\cZ_{N_0}/d\epsilon_0)\cdot\sigma^{-1}(\epsilon_0)=0$ implies 
$d\cZ_{N_0}/d\epsilon_0=0$, I calculate \cite{FC2}
\begin{equation}
\frac{d\log\cZ_{N_0}}{d \epsilon_0} = -C\epsilon_0^s
\left\{\log\left[1+e^{\beta(\epsilon_0-\mu)} \right]
- \frac{\beta}{C\epsilon_0^s}\right\} = 0 . \label{HA!}
\end{equation}
If $s>0$, $\log P$ has one and only one maximum at $\epsilon_0,N_0>0$, 
so for any macroscopic systems (i.e. large enough $C$) there will be a 
degenerate subsystem on the lowest energy levels at any temperature. 
If $s=0$ (e.g. particles in a two dimensional flat potential or in a 
one-dimensional harmonic potential) there is a transition temperature, 
$T_{c,2D}$, below which the maximum of $P(N_0)$ moves from $N_0=0$ to 
$N_0>0$, i.e. a degenerate gas forms. The degenerate gas may be put in 
correspondence with the 
the Bose-Einstein condensate in a gas of bosons with similar spectrum 
\cite{holthaus,FC1,FC2,FC3} and for simplicity I shall call it 
the {\em Fermi condensate} or the degenerate subsystem. The most 
interesting case seems to be $s<0$, 
when $\log P$ has either only one maximum, at $N_0=0$, or two maxima, 
at $N_0=0$ and $N_0>0$. As the temperature decreases, the second maximum 
increases and becomes bigger than the maximum at $N_0=0$. 

The rest of the article is organized as follows: in Section 2 I 
discuss in some detail the probability distribution $P_{N_0}(N_0)$ 
for the two-dimensional Fermi gas and calculate the relevant parameters 
of the Fermi condensate. In Section 3 I use in the standard way the 
grandcanonical formalism to calculate the thermodynamic quantities. 
In section 4 I compare the entropy obtained in section 3 to the entropy 
of the thermodynamically equivalent Bose gas and point out the miss-fit 
from the low temperatures range. The solution is suggested in 
section 5: in the canonical or microcanonical ensemble the condensate 
region should be considered separately and the Fermi distribution 
applies only in the {\em thermally active layer}, which is the energy 
interval above the condensate. 
The last section is reserved for conclusions. 

\section{Fermi condensation at constant density of states}\label{2D}

For constant $\sigma$, Eq. (\ref{HA!}) has a solution if and only if 
$\log[1+e^{-\beta\mu}]<(\sigma k_B T)^{-1}$. Therefore, as in \cite{FC2}, 
I define the condensation temperature $T_{c,F}$ by the equation 
\begin{equation} \label{T2cf}
\log[1+e^{-\beta_{c,F}\mu}]=(\sigma k_B T_{c,F})^{-1} \,.
\end{equation}
If I assume that in the temperature range of interest 
$\sigma k_B T\gg 1$ (i. e. I consider macroscopic systems), then 
also $\beta\mu \gg 1$, and the equation above may be written as 
\begin{equation}
N \approx \sigma k_B T_{c,F}\log(\sigma k_B T_{c,F}) \,. \label{T2cfN}
\end{equation}
In Eq. (\ref{T2cfN}) I used the approximation $N\approx \sigma\mu$, valid at 
low temperatures (see further, Eq. \ref{NwN0}). 
For $T<T_{c,F}$, 
$1\gg\log[1+e^{\beta(\epsilon_0-\mu)}]\approx e^{\beta(\epsilon_0-\mu)}$ and 
the solution of Eq. (\ref{HA!}) may be approximated by 
\begin{equation} \label{epsilon02d}
\epsilon_{0,max} = \mu -k_B T\log[\sigma k_B T] \, .
\end{equation}
The particle number that is associated to $\epsilon_{0,max}$ is 
$N_{0,max}=\epsilon_{0,max}\sigma = \sigma\mu-\sigma k_B T\log[\sigma k_B T]$. 
The distribution (\ref{logZN0}) is not symmetric and 
$\langle N_0\rangle < N_{0,max}$. In the low temperature limit, 
$\langle N_0\rangle$ converges to $N_{0,max}$, 
but general, closed analytical expressions for $\langle N_0\rangle$ seem 
difficult to find. 

Let me denote $\langle\epsilon_0\rangle \equiv\langle N_0\rangle/\sigma$. 
In the next sections we shall see that in a canonical {\em finite} 
system we have to consider that all the
energy levels from 0 to $\langle\epsilon_0\rangle\lesssim\epsilon_{0,max}$ 
are occupied, and the Fermi distribution applies only to the energy levels 
from $\langle\epsilon_0\rangle$ up-wards. 
For this reason, as mentioned in the 
Introduction, I shall say that the particles in the 
energy interval above $\langle\epsilon_0\rangle$ form the 
{\em thermally active layer}. Obviously, both $\langle\epsilon_0\rangle$ 
and $\langle N_0\rangle$ are subject to fluctuations. 
By comparing Eq. (\ref{epsilon02d}) or (\ref{HA!}) with Eq. (\ref{T2cf}), 
we observe that $T$ is the condensation temperature for the gas in the 
thermally active layer. 
The total number of particles in the system is then calculated as 
\begin{eqnarray}
N &=& \langle N_0\rangle + \int_{\langle\epsilon_0\rangle+\sigma^{-1}}^\infty
\frac{\sigma \, d\epsilon}{e^{\beta(\epsilon-\mu)}+1} = \sigma\mu-1 
\nonumber \\ 
&& +\sigma k_B T\log\left[1+e^{\beta(\langle\epsilon_0\rangle
+\sigma^{-1}-\mu)}\right]\approx \sigma\mu -1 \nonumber \\
&& +\sigma k_B Te^{\beta(\langle\epsilon_0\rangle+\sigma^{-1}-\mu)} \approx 
\sigma\mu +e^{\beta(\langle\epsilon_0\rangle-\epsilon_{0,max}+\sigma^{-1})} 
\nonumber \\
&&-1 \approx \sigma\mu +\beta(\langle\epsilon_0\rangle-\epsilon_{0,max}
+\sigma^{-1}) \nonumber \\
&\approx& \sigma\mu + \beta(\langle\epsilon_0\rangle-\epsilon_{0,max})+
e^{\beta(\epsilon_{0,max}-\mu)} \equiv \sigma\epsilon_F \,, 
\label{NN0}
\end{eqnarray}
where $\epsilon_F$ is the Fermi energy and I used the fact that 
$\sigma k_B Te^{\beta(\epsilon_{0,max}-\mu)}=1$ (Eq. \ref{HA!} or 
\ref{epsilon02d}). 

In the grandcanonical ensemble, if we do not take into account the 
degenerate subsystem, the total particle number is 
\begin{equation}
\tilde N = \sigma\mu +\sigma k_B T\log\left[1+e^{-\beta\mu}\right] \,.
\label{NwN0}
\end{equation}
The Fermi energy, denoted in this case by $\tilde\epsilon_F$, 
is given by the equation 
$\tilde N=\sigma\tilde\epsilon_F$, so for $\beta\mu\gg 1$ we have 
\begin{equation}
\tilde\epsilon_F \approx \mu+ k_B T e^{-\beta\mu} \,. \label{epsFt}
\end{equation}

\section{Grandcanonical Fermi gas}

Under grandcanonical conditions, the entropy and internal energy of 
a Fermi gas have the expressions:
\begin{eqnarray}
S &=& -\sigma k_B^2 T\left[2 Li_2(-e^{\beta\mu})+ 
\beta\mu\log(1+e^{\beta\mu}) \right] \label{Sgc} \\
{\rm and} && U = -(k_BT)^2\sigma Li_2(-e^{\beta\mu}) \,, \label{Ugc}
\end{eqnarray}
respectively. If we assume that different ensembles are equivalent, 
then we can use Eqs. (\ref{NwN0}) and (\ref{Ugc}) to express 
$\mu$ and $T$ in terms of $N$ and $U$ and plug these expressions into 
$S$. The result should be the same. Vice-versa, calculating 
$T=(\partial S/\partial U)^{-1}$ and $-\mu/T=(\partial S/\partial N)^{-1}$, 
one should obtain the grandcanonical temperature and chemical potential. 

\begin{figure}[t]
\begin{center}
\unitlength1mm\begin{picture}(60,55)(0,0)
\put(0,0){\psfig{file=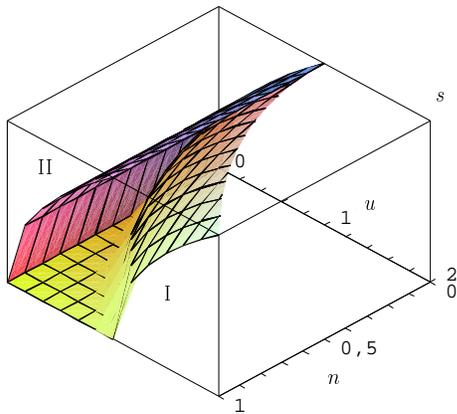,width=60mm}}
\end{picture}
\caption{Surface I: the entropy of a Fermi system, $s\equiv S/k_B\sigma$, 
as a function of $N\equiv N/\sigma$ and $u\equiv U/\sigma$. The range 
on the vertical axis is from 0 to $\pi\sqrt{2/3}$ (see Eq. \ref{Sapprox}). 
Surface II is obtained from surface I by substracting from the energy $U$, 
the ground state energy of the system, $U_{g.s.}(N)=N^2/2\sigma$.}
\label{entropy}
\end{center}
\end{figure}

To illustrate this change of variables, I shall express $S$ (Eq. \ref{Sgc}) 
in terms of $U$ and $N$ in the limit of low temperatures. 
In this limit $\beta\mu\gg 1$ and I can neglect $e^{-\beta\mu}$ from 
Eq. (\ref{NwN0}), retaining 
\begin{equation}
N \approx \beta\mu \,. \label{Napprox} 
\end{equation}
On the other hand, using the expansion 
\begin{equation}
Li_2(-e^{\beta\mu}) \approx -\frac{(\beta\mu)^2}{2}\left[1+
\frac{\pi^2}{3(\beta\mu)^2}\right] \label{Liexpansion}
\end{equation}
and Eq. (\ref{Napprox}), $U$ may be approximated as 
\begin{equation}
U \approx \frac{N^2}{2\sigma}\left[1+
\frac{\pi^2}{3}\cdot\left(\frac{\sigma k_B T}{N}\right)^2\right] \,.
\label{Uapprox} 
\end{equation}
Eliminating $k_BT$ and $\beta\mu$ from Eqs. (\ref{Napprox}) and 
(\ref{Uapprox}) plugging it into (\ref{Sgc}) I get 
\begin{equation}
\frac{S}{k_B\sigma} \approx \pi\sqrt{\frac{2}{3}\cdot\left(\frac{U}{\sigma} 
-\frac{N^2}{2\sigma^2}\right)} \,. \label{Sapprox}
\end{equation}
The result (\ref{Sapprox}) is plotted as surface I in Fig. \ref{entropy}. 

Since $U_0(N)\equiv N^2/2\sigma$ is the zero temperature energy of the 
Fermi gas, let me denote the excitation energy by $U_B\equiv U-U_0(N)$ 
and define $S_B(U_B,N)\equiv S(U_B+U_0(N),N)$. A gas that has 
the internal energy $U_B$ and entropy $S_B(U_B,N)$ as defined above, is 
called {\em thermodynamically equivalent} with the original Fermi gas 
(equivalence classes were defined in \cite{FC1}). The chemical potential 
of the gas $B$ is related to the chemical potential of the Fermi gas 
by $\mu_B=\mu-dU_0/dN=\mu-\epsilon_F$. In the low temperature 
limit 
\begin{equation}
\frac{S_B}{k_B\sigma} \approx \pi\sqrt{\frac{2}{3}\cdot\frac{U_B}{\sigma}} 
\,, \label{SBapprox}
\end{equation}
which is plotted as surface II in Fig. \ref{entropy}.
Another expression for $U_B$ may be obtained by using Landen's 
relation, 
$Li_2(-y)+Li_2[y/(y+1)]=-\frac{1}{2}\log^2(1+y)$ \cite{lewin,MHlee}: 
\begin{equation}
U_B = (k_B T)^\sigma Li_2[(1+e^{-\beta\mu})^{-1}] \,. 
\label{LandenU}
\end{equation}
Applying Landen's relation to Eq. (\ref{Sgc}), I get 
\begin{eqnarray}
S &=& -\sigma k_B^2 T\left[2 Li_2(-e^{\beta\mu})+ 
\beta\mu\log(1+e^{\beta\mu}) \right] \,. \label{SgcB} 
\end{eqnarray}

\section{The equivalent Bose gas}

It is well known that all the ideal gases of the same 
constant DOS and any exclusion statistics fall in the same equivalence 
class \cite{FC1,FC3,may,MHlee,apostol,crescimanno}. Since the ground state 
energy of a Bose gas (energy at zero temperature) is zero, $U_B$ defined 
in the previous section is 
nothing but the energy of the Bose gas and the surface 
II in Fig. \ref{entropy} is the entropy of the Bose gas. 
From the bosonic perspective, 
\begin{equation}
U_B = \int_0^\infty \frac{\sigma d\epsilon}{e^{\beta(\epsilon-\mu_B)}-1} 
= \sigma(k_BT)^2 Li_2(e^{\beta\mu_B}) \,. \label{UB}
\end{equation}
and from Eqs. (\ref{UB}) and (\ref{LandenU}) I obtain 
$e^{-\beta\mu_B}=1+e^{-\beta\mu}$

In principle a 
macroscopic Bose gas of constant DOS does not condense, but in a finite 
system the ground state becomes macroscopically populated below a 
``condensation'' temperature, $T_{c,B}$. The transition is not 
sharp, so $T_{c,B}$ may only be estimated. Grossmann and Holthaus 
\cite{holthaus} used the equation 
\begin{equation}
N = \sigma k_B T_{c,B} \log N \label{T2cbGH}
\end{equation}
as definition for $T_{c,B}$. Comparing $T_{c,B}$ with $T_{c,F}$, 
one can easily see that $T_{c,F}>T_{c,B}$, but they differ only by a 
factor 
$T_F/T_B=1+\log[\log(\sigma k_BT_{c,F})]/\log(\sigma k_BT_{c,F})\gtrsim 1$.

Let me now examine closer the equivalence between the Bose and Fermi 
gasses below $T_{c,F}$. At these temperatures, the ground state population 
of the Bose gas is 
\begin{equation}
N_0 = \frac{1}{e^{-\beta\mu_B}-1}\approx(-\beta\mu_B)^{-1}= 
e^{\beta\mu} \approx \sigma k_BT_{c,F} \,, \label{N0BC}
\end{equation}
which is of the order of $N$. In such a case the population of the 
ground state should be considered separately in all the equations. 
The total particle number in the Bose system should then be 
written as 
\begin{eqnarray}
N&=&N_0+\int_{\sigma^{-1}}^\infty \frac{\sigma \,d\epsilon}{
e^{\beta(\epsilon-\mu_B)}-1} = N_0 \nonumber \\
&& +\sigma k_BT \log\left[1-e^{\beta(\sigma^{-1}-\mu_B)} \right] \,. 
\label{NNOcond}
\end{eqnarray}
Much below $T_{c,F}$, $\sigma\mu_B\ll 1$ and Eq. (\ref{NNOcond}) 
may be approximated by 
\begin{equation}
N-N_0 \approx \sigma k_BT \log(\sigma k_BT) \,,
\end{equation}
which is equivalent to Eq. (\ref{epsilon02d}). 

Now a natural question arises: does the thermodynamic equivalence still 
holds at temperatures below condensation? For example at a temperature 
$T=T_{c,B}/2$, from Eq. (\ref{T2cbGH}) I get 
$\beta\mu\approx (2\sigma\mu/N)\log N\approx 2\log N$, 
which, if plugged into the expression (\ref{N0BC}) for $N_0$, gives 
\begin{equation}
N_0 \approx e^{2\log N} = N^2 \gg N \,! \label{hihi}
\end{equation}
One way to interpret this is that the thermodynamic equivalence breaks 
down at temperatures around and below $T_{c,F}$, i.e. when the population 
of the ground state becomes macroscopic. Therefore, below $T_{c,F}$ 
the transformation $U\to U_B$ and $S\to S_B$, presented in the 
previous section, does not lead to the bosonic entropy. But this 
cannot be true, since it was shown that there exist a one-to-one 
correspondence between the microscopical configurations of a Bose and 
a Fermi system, of identical constant DOS and the same excitation energy 
\cite{FC1,crescimanno}. 
Therefore, the canonical entropies and partition functions ought to be 
the same. In this case, what do we have to tailor? The answer is: 
{\em the canonical partition function of the Fermi gas}. 

\section{Canonical Fermi gas}

By the exclusion statistics transformation (EST) method \cite{FC3,FC1} 
every distribution of fermions along the single particle energy axis 
is transformed into a distribution of bosons. The $N_0$ fermions 
that form the Fermi condensate are mapped onto bosons on the 
ground state (the Bose condensate). For a system of constant DOS the 
rest of the particles have a Fermi distribution and are mapped onto 
a Bose distribution of particles. This transformation have been made 
explicitly in Section V.A of Ref. \cite{FC3}. The bosonic chemical potential 
is given by Eq. (\ref{NNOcond}), with 
$N_0\equiv \langle N_0\rangle=(e^{-\beta\mu_B}-1)^{-1}$. This fixes also 
the fermionic chemical potential to $\mu=\epsilon_F+\mu_B$ and 
the number of fermions in the condensate to $\langle N_0\rangle$. 
As a consequence of this, in the 
limit of low temperatures $\beta(\epsilon_F-\mu)$ does not have 
the expected asymptotic behavior 
$\beta(\epsilon_F-\mu)\approx e^{-\beta\epsilon_F}$, but instead, from 
Eq. (\ref{NNOcond}) I get 
\begin{eqnarray}
\beta(\epsilon_F-\mu) &=& \beta\mu_B = 
[N-\sigma k_B T\log(\sigma k_B T)]^{-1} \nonumber \\
&\approx& \frac{1}{N}+\frac{\sigma k_B T\log(\sigma k_B T)}{N^2} \,.
\label{last}
\end{eqnarray}

In addition, Jaynes' theory -- according to which the probability 
associated to each microscopic configuration should have the form 
(\ref{psEN}) -- seems to contain intrinsic contradictions. 
The distribution (\ref{psEN}) represents the least biased estimate 
given that the average number of particles in the system is $N$ and 
the average energy is $U$. On the other hand, the bosonic distribution 
of the equivalent Bose gas is the least biased estimate, given the 
average {\em excitation energy} $U_B$ and particle number $N$. As 
shown above, the two distributions {\em do not map onto each-other over the 
whole spectrum}. In Ref. \cite{FC3} it is proven that the situation is 
even more dramatic when the DOS is not constant. 

\section{Conclusions}

In conclusion I discussed the effect of fermionic condensation -- 
which is the apparition of a degenerate subsystem at the bottom of the 
single particle spectrum -- in a system of constant density of states. 
This leads to a correction in the calculation of the chemical potential 
in the canonical ensemble at low temperatures (see Eq. \ref{last}), 
which further implies that the canonical and grandcanonical ensembles 
are not equivalent at low temperatures even for Fermi systems. 

By applying the exclusion statistics transformation to the 
Fermi system, one obtains a (thermodynamically equivalent \cite{FC1}) 
Bose system. If the Fermi system is 
condensed, the degenerate subsystem is mapped onto the Bose-Einstein 
condensate -- from where the name of Fermi condensate is derived. 
The condensate 
fluctuates and the value of these fluctuations are simply given by 
the condensate fluctuations of the equivalent Bose system. 
For the calculation of Bose ground state fluctuations in 
canonical end microcanonical ensembles I refer 
the reader to the articles of Holthaus et al. (e.g. \cite{holthaus,holthaus1} 
and citations therein) and Tran et al. (e.g. \cite{tran} and 
citations therein). 

Non-equivalence between grandcanonical Bose and Fermi 
gases have also been observed very recently by Patton et al. \cite{patton} 
in computer simulations of small systems.


\begin{thebibliography}{10}

\bibitem{jaynes} E. T. Jaynes, Phys. Rev. {\bf 106}, 620 (1957);  Phys. Rev. {\bf 108}, 171 (1957). 
\bibitem{lee}  K. Lee, Phys. Rev. E {\bf 53}, 6558 (1996); J. Lee and K.-C. Lee, Phys. Rev. E {\bf 62}, 4558 (2000).
\bibitem{FC1} D. V. Anghel, J. Phys. A: Math. Gen. {\bf 35}, 7255-7267 (2002), cond-mat/0105089.
\bibitem{FC2} D. V. Anghel, J. Phys. A: Math. Gen. {\bf 36}, L577(2003); cond-mat/0310248.
\bibitem{FC3} D. V. Anghel, submitted to J. Math. Phys.; cond-mat/0310377.
\bibitem{holthaus} S. Grossmann and M. Holthaus, Phys. Rev. E {\bf 54}, 3495 (1996).
\bibitem{jin} B. DeMarco, S. B. Papp, and D. S. Jin, Phys. Rev. Lett. {\bf 86}, 5409 (2001).
\bibitem{may} R. M. May, Phys. Rev. {\bf 135}, A1515 (1964).
\bibitem{MHlee} M. H. Lee, Phys. Rev. E {\bf 55}, 1518-1520 (1997).
\bibitem{apostol} M. Apostol, Phys. Rev. E {\bf 56}, 4854 (1997).
\bibitem{crescimanno} M. Crescimanno, and A. S. Landsberg, Phys. Rev. A {\bf 63}, 35601 (2001); cond-mat/0003020.
\bibitem{lewin} L. Lewin, {\em Dilogarithms and associated functions} (McDonald, London, 1958).
\bibitem{lee1} M. H. Lee, J. Math. Phys. {\bf 36}, 1217 (1995).
\bibitem{holthaus1} M. Holthaus and E. Kalinowski, Ann. Phys. (N.Y.) {\bf 276}, 321 (1999).
\bibitem{tran} M. N. Tran, J. Phys. A, Math. Gen. {\bf 36}, 961 (2003). 
\bibitem{patton} K. R. Patton, M. R. Geller, and M. P. Blencowe, cond-mat/0406285.

\end{thebibliography}
\end{document}